\documentclass{amsart}

\usepackage{amssymb,latexsym,amsmath,amsfonts,hhline,comment,mathrsfs,stmaryrd}
\usepackage{pdfsync}
\usepackage{amsthm}
\usepackage{graphicx}
\usepackage{wrapfig}

\usepackage{color,soul}

\usepackage{enumerate}
\usepackage{tikz}

\usepackage{ifpdf}
\ifpdf \usepackage[colorlinks=true, citecolor=blue, linkcolor=blue, urlcolor=blue]{hyperref} \fi

\newtheorem{formula}{}[section]
\newtheorem{definition}[formula]{Definition}
\newtheorem{corollary}[formula]{Corollary}
\newtheorem{remark}[formula]{Remark}
\newtheorem{lemma}[formula]{Lemma}
\newtheorem{proposition}[formula]{Proposition}
\newtheorem{theorem}[formula]{Theorem}

\def\thrm{\begin{theorem}}
\def\thrml#1{\begin{theorem}\label{#1}}
\def\ethrm{\end{theorem}}
\def\prpstn{\begin{proposition}}
\def\prpstnl#1{\begin{proposition}\label{#1}}
\def\eprpstn{\end{proposition}}
\def\rmrk{\begin{remark}}
\def\rmrkl#1{\begin{remark}\label{#1}}
\def\ermrk{\end{remark}}
\def\dfntn{\begin{definition}}
\def\dfntnl#1{\begin{definition}\label{#1}}
\def\edfntn{\end{definition}}
\def\nmrt{\begin{enumerate}}
\def\enmrt{\end{enumerate}}
\def\tm#1{\item[{\rm (#1)}]}
\def\qtnl#1{\begin{equation}\label{#1}}
\def\eqtn{\end{equation}}
\def\lmm{\begin{lemma}}
\def\lmml#1{\begin{lemma}\label{#1}}
\def\elmm{\end{lemma}}
\def\crllr{\begin{corollary}}
\def\crllrl#1{\begin{corollary}\label{#1}}
\def\ecrllr{\end{corollary}}
\def\css{\begin{cases}}
\def\ecss{\end{cases}}
					
\def\proof{\noindent{\bf Proof}.\ }
\def\eprf{\hfill$\square$}

\def\cE{{\mathcal E}}

\def\cH{{\mathcal H}}

\def\cT{{\mathcal T}}

\def\fK{{\frak K}}

\DeclareMathOperator{\aut}{Aut}

\DeclareMathOperator{\con}{Conn}

\DeclareMathOperator{\id}{id}

\DeclareMathOperator{\im}{Im}

\DeclareMathOperator{\iso}{Iso}

\DeclareMathOperator{\sym}{Sym}

\DeclareMathOperator{\poly}{poly}

\DeclareMathOperator{\WL}{WL}

\def\grp#1{\langle {#1}\rangle}
\def\ov{\overline}
\def\wh{\widehat}
\def\wt{\widetilde}
\def\qaq{\quad\text{and}\quad}

\def\phmb#1{{\phantom{x}\hspace{-2mm}^{#1}}}

\makeatletter 
\def\@seccntformat#1{\csname the#1\endcsname. } 
\def\@biblabel#1{#1.} 

\newcommand{\order}[1]{order-{${#1}$}}

\makeatother

\makeatletter 
\def\@seccntformat#1{\csname the#1\endcsname. } 
\def\@biblabel#1{#1.}

\newcommand{\E}{\mathcal{E}}

\title[Testing isomorphism of chordal graphs of bounded leafage]{Testing isomorphism of chordal graphs of bounded leafage is fixed-parameter tractable}
\author{Vikraman Arvind}
\address{The Institute of Mathematical Sciences (HBNI), Chennai, India}
\email{arvind@imsc.res.in}
\author{Roman Nedela}
\address{Faculty of Applied Sciences, University of West Bohemia, Technick\'a 8, Pilsen, Czech Republic}
\email{nedela@savbb.sk}
\author{Ilia Ponomarenko}
\address{V. A. Steklov Institue of Mathematics, Russian Academy of Sciences, St. Petersburg, Russia}
\email{inp@pdmi.ras.ru} 
\author{Peter Zeman}
\address{Department of Applied Mathematics, Faculty of Mathematics and Physics, Charles University, Prague, Czech Republic}
\email{zeman@kam.mff.cuni.cz}

\date{}

\begin{document}

\begin{abstract}
The computational complexity of the graph isomorphism problem is
considered to be a major open problem in theoretical computer science.
It is known that testing isomorphism of chordal graphs is
polynomial-time equivalent to the general graph isomorphism
problem. Every chordal graph can be represented as the intersection
graph of some subtrees of a representing tree, and the leafage of a
chordal graph is defined to be the minimum number of leaves in a
representing tree for it. We prove that chordal graph isomorphism is
fixed parameter tractable with leafage as parameter. In the process we
introduce the problem of isomorphism testing for higher-order
hypergraphs and show that finding the automorphism group of \order{k}
hypergraphs with vertex color classes of size $b$ is fixed parameter
tractable for any constant $k$ and $b$ as fixed parameter.
  
\end{abstract}

\maketitle

\section{Introduction}

The graph isomorphism problem is one of the few natural problems in NP
that is neither known to be NP-complete nor it is known to be
polynomial-time solvable.  In a fairly recent breakthrough,
Babai~\cite{babai_quasipolynomial} proved that the graph isomorphism
problem is solvable in quasipolynomial time, i.e., in time
$n^{\poly(\log n)}$, where $n$ is the number of vertices.\smallskip

A significant line of research concerns the parameterized complexity
of the graph isomorphism problem with respect to some natural graph
parameter. These include treewidth~\cite{fpt_treewidth},
degree~\cite{luks_degree,degree_improved},
genus~\cite{ponomarenko_genus,hypergraph_iso_neuen}, excluded
minors~\cite{ponomarenko_minors,minors_improved}, etc. It is worth
mentioning that in several of these cases, Babai's new techniques have
yielded new algorithms with improved running time. For example, Luks's
original algorithm with running time $n^{O(k)}$ for degree-$k$ graphs
has a modified $n^{\poly(\log k)}$-time
algorithm~\cite{luks_degree,degree_improved}. However, in some of
these cases a fixed-parameter tractable (FPT) algorithm, i.e., an
algorithm with running time $f(k)\poly(n)$, have remained elusive.
Such an improvement likely cannot be obtained using known techniques
and would require some new techniques and ideas.

\smallskip

In our work, we deal with parameterized complexity of the graph
isomorphism problem for the class of chordal graphs. An undirected
graph is said to be \emph{chordal} if it has no chordless cycle of
length at least four. Every chordal graph admits a representation as
the intersection graph of subtrees of some tree
$T$~\cite{gavril_chordal}. We say that a chordal graph $X$ has
representing tree $T$ if $X$ can be represented as the intersection
graph of subtrees of $T$. The \emph{leafage}~$ \ell(X)$ of a chordal
graph $X$ is the least positive integer such that $X$ has a
representing tree with $\ell(X)$ leaves. The notion of leafage was
introduced in~\cite{leafage} and is a natural graph parameter for
chordal graphs.

It is interesting to note that the well-studied interval graphs are
precisely the intersection graphs of paths. It follows that
$\ell(X)\le 2$ if and only if $X$ is an interval graph (and $\ell(X) =
1$ if and only if $X$ is complete). Thus, the leafage of a chordal
graph $X$ measures how far it is from being an interval graph, which
has interesting algorithmic consequences.  For instance, efficient
solutions to certain NP-hard problems on interval graphs naturally
extend to chordal graphs of bounded leafage;
e.g,~\cite{stacho_subcoloring}.\smallskip

Graph Isomorphism restricted to chordal graphs is polynomial-time
equivalent to Graph Isomorphism for general
graphs~\cite[Theorem~5]{linear_interval_isomorphism}. On the other
hand, the problem can be solved in polynomial (even linear time) for
interval graphs~\cite{linear_interval_isomorphism}. The main result of
the present paper can be considered as a substantial generalization of
the latter.\smallskip

\noindent\textbf{Results of this paper}\smallskip

\thrml{010721w}
Testing isomorphism of chordal graphs of leafage $\ell$ is fixed
parameter tractable, with $\ell$ as fixed parameter.
\ethrm

The leafage of chordal graphs is known to be polynomial-time
computable \cite{HS09}. Denote by $\fK_\ell$ the class of all chordal
graphs of leafage at most $\ell$. In particular, the graph class
$\fK_\ell$ is polynomial-time recognizable.

In order to test if two connected graphs $X,Y\in\fK_\ell$ are
isomorphic, it suffices to check if there is a generator of the
automorphism group of their disjoint union $X\cup Y$, which swaps $X$
and $Y$. Since the graph $X\cup Y$ belongs to the class $\fK_{2\ell}$,
the graph isomorphism problem for the graphs in $\fK_{\ell}$ is
reduced to the problem of determining the automorphism group of a
given graph in $\fK_{2\ell}$. Thus Theorem~\ref{010721w} is an
immediate consequence of the following theorem which is proved in the
paper.

\thrml{180221f}
Given an $n$-vertex graph $X\in\fK_\ell$, a generating set of the group $\aut(X)$ can
be found in time $t(\ell)\poly(n)$, where  $t(\cdot)$ is a  function independent of~$n$.
\ethrm

The function $t$ from Theorem~\ref{180221f} is bounded from above by a
polynomial in~$(\ell 2^\ell)!$. The running time bound, especially the
function $t$, does not appear to be final and, most likely, it can be
significantly improved. \smallskip

We emphasize that our algorithm does not require that the input $X$ is
given by an intersection representation. Indeed, the algorithm works
correctly on all chordal graphs and the leafage bound $\ell$ is
required only to bound the running time for inputs from the class
$\fK_\ell$.

The proof of Theorem~\ref{180221f} is given in
Section~\ref{010721z}. The main steps involved in the algorithm are:
(a) to efficiently transform the given graph $X$ into an \order{3}
hypergraph $H=H(X)$ (see below), (b) to give an algorithm for
computing a generating set for $\aut(H)$, and (c) to recover from it a
generating set for $\aut(X)$.


This brings us to the notion of \emph{higher-order hypergraphs}. A
usual hypergraph with vertex set $V$ has hyperedge set contained in
the power set $\cE_1=2^V$. The hyperedges of an \order{3} hypergraph
$H$ will, in general, include \order{2} and \order{3}
hyperedges. These are elements of $\cE_2=2^{\cE_1}$ and
$\cE_3=2^{\cE_2}$, respectively. The hyperedge set $E$ of $H$ is
contained in $\cE_1\cup \cE_2\cup \cE_3$ and can be of
triple-exponential size in $|V|$. However, the input size of $H$ is
defined to be $|V|\cdot |E|$, for $H$ given as input to an
algorithm. The efficient reduction from finding $\aut(X)$ to finding
$\aut(H)$ is presented in Sections~\ref{180221z}
and~\ref{250621q}. The key point of the reduction is a
graph-theoretical analysis of the vertex coloring of the chordal
graph~$X$ obtained by the $2$-dimensional Weisfeiler-Leman
algorithm~\cite{Weisfeiler1968}. The reduction takes $X$ as input and
computes the colored \order{3} hypergraph $H$ such that each vertex
color class of $H$ has size at most $b=\ell 2^\ell$, where
$\ell=\ell(X)$.
\smallskip

At this point, we deal with the general problem of determining the
automorphism group of a colored \order{k} hypergraph $H$ ($k\ge 1)$ by
an FPT algorithm with respect to the parameter~$b$ which bounds the
size of each vertex color class. This problem seems interesting in
itself and could find other applications. For ordinary hypergraphs,
it was shown to be fixed parameter tractable
in~\cite{colored_hypergraph_fpt}. A generalization of that result to
\order{k} hypergraphs is given in Section~\ref{250621y}. The running
time bound we obtain is not FPT in terms of the parameter $k$.
Whether or not the problem is in FPT with both $b$ and $k$ as fixed
parameters seems to be an interesting problem.\smallskip

We complete the introduction with some remarks about $H$-graphs
introduced in~\cite{precoloring}. An $H$-graph $X$ is an intersection
graph of connected subgraphs of a subdivision of a fixed graph $H$.
Every graph is an $H$-graph for a suitable $H$, which gives a
parametrization for all graphs. It is interesting to note that we can
get well-known graph classes as $H$-graphs for suitable choices of
$H$. For instance, interval graphs are $K_2$-graphs, circular-arc
graphs are $K_3$-graphs, and chordal graphs are the union of all
$T$-graphs, where $T$ is a tree.\smallskip

Basic algorithmic questions on $H$-graphs, including their recognition
and isomorphism testing, have been studied,
e.g.,~\cite{h-graphs,h-graphs_comb,h-graphs_opt}.  It is shown
in~\cite{sd-graphs} that isomorphism testing for $S_d$-graphs, where
$S_d$ is a star of degree $d$, is fixed parameter tractable. Since
$S_d$-graphs are chordal graphs of leafage at most~$d$, our FPT
algorithm applied to chordal graphs with bounded leafage significantly
extends that result~\cite{sd-graphs}.\footnote{Paper \cite{AH21}
  appeared in the arXiv some time after our paper was published
  there, contains an FTP algorithm testing isomorphism of T-graphs for
  every fixed tree T.  This result gives an alternative FPT algorithm
  for chordal graphs of leafage $\ell$.}  On the other hand, the
isomorphism problem for $H$-graphs is as hard as the general graph
isomorphism problem if $H$ is not unicyclic~\cite{h-graphs_gic}.
Thus, it remains open whether isomorphism can be solved in polynomial
time for the unicyclic case with fixed number of leaves, which would
provide a dichotomy for the parametrization by $H$-graphs.  Our work
can be also considered a step towards this dichotomy.

\section{Preliminaries}
 
\subsection{General notation}

Throughout the paper, $\Omega$ is a finite set. Given a bijection~$f$
from $\Omega$ to another set and a subset $\Delta\subseteq\Omega$, we
denote by $f^\Delta$ the bijection from~$\Delta$ to its image
$\Delta^f=\{\delta^f:\ \delta\in\Delta\}$. For a set $S$ of
bijections from $\Delta$ to another set, we put
$S^\Delta=\{f^\Delta:\ f\in S\}$.\smallskip
 
The group of all permutations of a set $\Omega$ is denoted by
$\sym(\Omega)$. When a group~$G$ acts on $\Omega$, we set
$G^\Omega=\{g^\Omega:\ g\in G\}$ to be the permutation group induced
by this action. Concerning standard permutation group
algorithms we refer the reader to~\cite{Seress}.\smallskip
 
Let $\pi$ be a partition of $\Omega$. The set of all
unions of the classes of~$\pi$ is denoted by~$\pi^\cup$. The partition
$\pi$ is a \emph{refinement} of a partition $\pi'$ of $\Omega$ if each
class of $\pi'$ belongs to~$\pi^\cup$; in this case, we write $\pi\ge
\pi'$, and $\pi>\pi'$ if $\pi\ge \pi'$ and $\pi\ne\pi'$. The partition
of $\Delta\subseteq\Omega$ induced by $\pi$ is denoted by
$\pi_\Delta$.
 
\subsection{Graphs}
Let $X$ be an undirected graph. The vertex and edge sets of $X$ are denoted
by~$\Omega(X)$ and $E(X)$, respectively.  The automorphism group of
$X$ is denoted by $\aut(X)$. The set of all isomorphisms from $X$ to a
graph $X'$ is denoted by $\iso(X,X')$.\smallskip

The set of all leaves and of all connected components of $X$
are denoted by~$L(X)$ and $\con(X)$, respectively. For a vertex
$\alpha$, we denote by~$\alpha X$ the set of neighbors of $\alpha$ in
$X$. The vertices $\alpha$ and $\beta$ are called \emph{twins} in $X$
if every vertex other than $\alpha$ and $\beta$ is adjacent
either to both $\alpha$ and $\beta$ or neither of them. The graph $X$
is said to be {\it twinless} if no two distinct vertices of $X$ are
twins.\smallskip

Let $\Delta,\Gamma\subseteq \Omega(X)$. We denote by
$X_{\Delta,\Gamma}$ the graph with vertex set $\Delta\cup\Gamma$ in
which two vertices are adjacent if and only if one of them is in
$\Delta$, the other one is in $\Gamma$, and they are adjacent in
$X$. Thus, $X_\Delta=X_{\Delta,\Delta}$ is the subgraph of $X$ induced
by~$\Delta$, and $X_{\Delta,\Gamma}$ is bipartite if
$\Delta\cap\Gamma=\varnothing$. \smallskip

Let $\Delta\subseteq \Omega(X)$ and $Y=X_\Delta$. The set of all
vertices adjacent to at least one vertex of~$\Delta$ and not belonging
to $\Delta$ is denoted by $\partial Y$.  The subgraph of $X$, induced
by $\Delta\cup \partial Y$ is denoted by $\ov Y$.\smallskip

For a tree $T$, let $S(T) = \{\Omega(T') : T' \text{ is a subtree of }
T\}$ be the set of all vertex sets of the subtrees of $T$. A
\emph{representation} of a graph $X = (\Omega, E)$ on the tree $T$
(called \emph{tree-representation}) is a function $R\colon \Omega \to S(T)$
such that for all $u, v \in \Omega$,
$$
R(u) \cap R(v) \ne \varnothing\Leftrightarrow \{u,v\}\in E.
$$

It is known that a graph~$X$ is chordal if and only if $X$ has a
tree-representation~\cite{gavril_chordal}. The leafage $\ell(X)$
of~$X$ is defined to be the minimum of $|L(T)|$ over all trees $T$
such that $X$ has a tree-representation on $T$.


\subsection{Colorings}\label{180721a}
A partition $\pi$ of $\Omega$ is said to be a \emph{coloring} (of
$\Omega$) if the classes of $\pi$ are indexed by elements of some set,
called \emph{colors}. In this case, the classes of~$\pi$ are called
\emph{color classes} and the color class containing $\alpha\in\Omega$
is denoted by $\pi(\alpha)$. Usually the colors are assumed to be
linearly ordered. A bijection $f$ from $\Omega$ to another set
equipped with coloring $\pi'$ is said to be \emph{color preserving} if
the colors of $\pi(\alpha)$ and $\pi'(f(\alpha))$ are the same for all
points $\alpha\in \Omega$.


A graph equipped with a coloring of the vertex set (respectively, edge
set) is said to be \emph{vertex colored} (respectively, \emph{edge
  colored}); a graph that is both vertex and edge colored is said to
be \emph{colored}. The isomorphisms of vertex/edge colored graphs are
ordinary isomorphisms that are color preserving. To emphasize this, we
sometimes write $\aut(X,\pi)$ for the automorphism group of a
graph~$X$ with coloring~$\pi$.\smallskip

Let $X$ be a colored graph with vertex coloring $\pi$. Consider the
application of the Weisfeiler-Leman algorithm ($2$-dim WL) to
$X$~\cite{Weisfeiler1968}. For the purpose of the paper, it suffices
to understand that $2$-dim WL iteratively colors pairs of vertices of
$X$ until the coloring satisfies a specific regularity condition
(where the vertex coloring corresponds to the coloring of diagonal
pairs $(\alpha,\alpha)$). The resulting coloring of pairs is just what
is called a \emph{coherent configuration}.

The output of $2$-dim WL defines a new vertex coloring
$\WL(X,\pi)\ge\pi$ of $X$. We say that $\pi$ is \emph{stable} if
$\WL(X,\pi)=\pi$. In the language of coherent configurations, $\pi$ is
stable precisely when the classes of $\pi$ are the fibers of a
coherent configuration (details can be found in the
monograph~\cite{CCBook}). In the sequel, we will use some elementary
facts from theory of coherent configurations. The following statement
summarizes relevant properties of stable colorings.

\lmml{230621a}\label{stable-lem}
Let $X$ be a graph and $\pi$ be a stable coloring of $X$. Then
\nmrt \tm{1} for $\Delta,\Gamma\in\pi$, the number $|\delta X\cap
\Gamma|$ does not depend on $\delta\in \Delta$, \tm{2} if
$\Delta\in\pi^\cup$ or $X_\Delta\in\con(X)$, then the
coloring $\pi_\Delta$ is stable.
\enmrt
\elmm

A coloring $\pi$ of the vertices of a graph $X$ is said to be
\emph{invariant} if every class of~$\pi$ is $\aut(X)$-invariant. In
this case, the coloring $\WL(X,\pi)$ is also invariant and
stable. Since the coloring of the vertices in one color is invariant
and the Weisfeiler-Leman algorithm is polynomial-time, in what follows
we deal with invariant stable colorings.

\subsection{Hypergraphs}\label{hohg-defn}
Let $V$ be a finite set. The set $\cE_k=\cE_k(V)$ of the
\emph{\order{k} hyperedges} on $V$ is defined recursively as follows:
$$
\E_0 = V,\qquad \E_k = \E_{k-1}\cup 2^{\E_{k-1}}\ \text{ for } k>1.
$$
So, we consider elements of~$V$ as \order{0} hyperedges and the
\order{k} hyperedges include all \order{(k-1)} hyperedges and their
subsets. \smallskip

Let $U\subseteq V$ and $e\in\cE_k$ ($k\ge 1)$. We recursively define
the {\it projection} of $e$ on $U$ as the multiset
$$
e^U=\css
e\cap U &\text{if $k=1$,}\\
\{\{\wt e\phmb{\,U}:\ \wt e\in e\}\} &\text{if $k>1$.}\\
\ecss
$$
We extend this definition to all sets $E\subseteq \cE_k$ by putting $E^U=\{e^U:\ e\in E\}$. 

\begin{definition}[\order{k} hypergraph]
An \order{k} hypergraph ($k\ge 1$) on $V$ is a pair $H=(V,E)$, where
$E\subseteq 2^{\E_k}$; the elements of $V$ and $E$ are called vertices
and hyperedges of~$H$, respectively.
\end{definition}

Clearly, \order{1} hypergraphs are usual hypergraphs. Moreover,
higher-order hypergraphs (i.e., \order{k} hypergraph for some $k$) are
combinatorial objects in the sense of ~\cite{brand_comb_obj}. The
concepts of isomorphism and coloring extend to higher-order
hypergraphs in a natural way. \medskip

Let $k\ge 2$. The {\it $(k-1)$-skeleton} of an \order{k} hypergraph
$H=(V,E)$ is an \order{(k-1)} hypergraph $H^{(k-1)}$ on~$V$ with the hyperedge set
$$
E^{(k-1)}=\{\wt e\in\cE_{k-1}:\ \wt e\text{ is an element of some } e\in\cE_k\cap E\}.
$$
It is easily seen that for every \order{k} hypergraph $H'=(V',E')$ 
\qtnl{110721a}
\iso(H,H')=\{f\in \iso(H^{(k-1)},{H'}\phmb{(k-1)}):\ e\in E^{(k)}\ \Leftrightarrow\ e^f\in E'^{(k)}\}.
\eqtn
where for each \order{k} hyperedge $e=\{e_1,\ldots,e_a\}$ we set $e^f=\{e_1^f,\ldots,e_a^f\}$.\smallskip

Let $H_1=(V_1,E_1)$ be an \order{k} hypergraph for some
  $k$ and $H_2=(V_2,E_2)$ be a usual hypergraph such that $V_2=E_1$.
Then each hyperedge $e\in E_2$ is a subset of hyperedges of $H_1$. We
define the \emph{hypergraph composition} of $H_1$ and $H_2$ to be the
\order{(k+1)} hypergraph
$$
H:=H_1\uparrow H_2=(V,E_1\cup E_2).
$$

When the hypergraphs $H_1$ and $H_2$ are colored, the vertex coloring
of $H$ is defined in the obvious way. The color $c(e)$ of $e\in E(H)$
is defined as follows: if $e\in E_1\setminus E_2$ then $c(e)$ is the
color $c_1(e)$ of $e$ in $H_1$. If $e\in E_1\cap E_2$ then $c(e)$ is
defined as the triple $(0,c_1(e),c_2(e))$, where $c_2(e)$ is the color
of $e$ in $H_2$. Finally, if $e\in E_2\setminus E_1$ then
$c(e)=(1,c_1(e'),c_2(e))$, where $e'$ is the set of elements of $e$.


\section{Chordal graphs}

\subsection{Stable colorings in chordal graphs}

In this subsection, we prove several auxiliary statements about the
structure of subgraphs of a chordal graph, induced by one or two color
classes of a stable coloring.

\lmml{230621a4} Let $X$ be a chordal graph and $\pi$ a stable coloring
of $X$. Then for every $\Delta,\Gamma\in\pi$, the following statements
hold: \nmrt \tm{1} $\con(X_\Delta)$ consists of cliques of the same
size, \tm{2} if $|\con(X_\Delta)|\le |\con(X_\Gamma)|$, then
$\con(X_\Delta)=\{Y_\Delta:\ Y\in \con(X_{\Delta\cup\Gamma})\}$,
\tm{3} if the graphs $X_\Delta$ and $X_\Gamma$ are complete, then
$X_{\Delta,\Gamma}$ is either complete bipartite or empty.  \enmrt
\elmm

\proof (1) By Lemma~\ref{230621a}(1) for $\Delta=\Gamma$, the graph
$X_\Delta$ is regular. It suffices to verify that every graph
$Y\in\con(X_\Delta)$ is complete. Since $Y$ is chordal it contains a
simplicial vertex, i.e., a vertex whose neighborhood induces a
complete graph. As~$Y$ is regular, all its vertices are
simplicial. Thus, $Y$ is complete.\smallskip

(2) Let $X'$ be a bipartite graph with parts $\Delta'=\con(X_\Delta)$
and $\Gamma'=\con(X_\Gamma)$ in which two vertices $\alpha'\in
\Delta'$ and $\beta'\in\Gamma'$ are adjacent if and only if there are
vertices $\alpha\in \alpha'$ and $\beta\in\beta'$ adjacent in~$X$. By
statement~(1), the components of $X_{\Delta\cup\Gamma}$ are in
one-to-one correspondence with the components of $X'$. Denote by $Y'$
the component of $X'$, corresponding to the component $Y\in
\con(X_{\Delta\cup\Gamma})$.


The partition $\pi'=\{\Delta',\Gamma'\}$ is a stable coloring of
$X'$. Indeed, it suffices to find a coherent configuration on
$\Omega(X')$, for which $\Delta'$ and $\Gamma'$ are fibers. As such a
configuration, one can take the quotient of the coherent configuration
corresponding to $\pi$, modulo the equivalence relation on
$\Delta\cup\Gamma$ the classes of which are vertex sets of the graphs
belonging to $\Delta'$ and $\Gamma'$, see~\cite[Section~3.1.2]{CCBook}.

By Lemma~\ref{230621a}(1), any two vertices of $X'$ that are from the
same part have the same degree. Moreover, the graph $X'$ is obviously
chordal. Consequently, it is acyclic: otherwise, $X'$ being bipartite
contains an induced cycle of length at least~$4$, which is impossible
for a chordal graph. Hence, $X'$ has a vertex $\alpha'$ of
degree~$1$. Since all vertices of the part containing $\alpha'$ have
the same degree, each $Y'\in\con(X')$ is a star. The center of this
star lies in $\Delta'$, because $|\Delta'|\le |\Gamma'|$.  Thus,
$Y_\Delta\in\con(X_\Delta)$, which implies the required
statement.\smallskip

(3) Without loss of generality we may assume that the bipartite graph
$X_{\Delta,\Gamma}$ is not empty and $\Delta\ne\Gamma$. Suppose to the
contrary that there are $\delta_1,\delta_2\in\Delta$ such that
$\delta_1 X\cap\Gamma\ne \delta_2 X\cap\Gamma$. By
Lemma~\ref{230621a}(1), we have $|\delta_1 X\cap\Gamma|=|\delta_2
X\cap\Gamma|$.  Thus there exist $\gamma_1\in \delta_1 X\cap\Gamma$
and $\gamma_2\in\delta_2 X\cap\Gamma$ such that
$$
\gamma_1\not\in \delta_2 X\quad\text{and}\quad \gamma_2\not\in \delta_1 X.
$$

By assumption, the graphs $X_\Delta$ and $X_\Gamma$ are complete.
Hence, $\delta_1$ and $\delta_2$ are adjacent, and also $\gamma_1$ and
$\gamma_2$ are adjacent.  Therefore, the vertices
$\delta_1,\gamma_1,\gamma_2,\delta_2$ form an induced $4$-cycle of
$X$, which is a contradiction. Thus any two vertices that are in
$\Delta$ have the same neighborhoods in the bipartite graph
$X_{\Delta,\Gamma}$. As $X_{\Delta,\Gamma}$ has no isolated vertices
in $\Gamma$ (by Lemma~\ref{stable-lem}), it follows that
$X_{\Delta,\Gamma}$ is a complete bipartite graph.
\eprf

\begin{remark}
Recall that stable colorings are defined via the $2$-dimensional
Weisfeiler-Leman algorithm. While the $1$-dimensional Weisfeiler-Leman
algorithm suffices for the first and third parts of
Lemma~\ref{230621a4}, it is worth noting that the second part requires
the $2$-dimensional algorithm.
\end{remark}

\lmml{180621c}
Let $X$ be a connected chordal graph and let $\pi$ be a stable partition of~$\Omega$.
There exists $\Delta \in \pi$ such that the graph $X_\Delta$ is complete.
\elmm

\proof The statement immediately follows from Lemma~\ref{230621a4}(1)
if $|\pi|=1$. Assume that $|\pi|>1$. Since $\pi$ is stable, the
classes of $\pi$ are the fibers of some coherent configuration
on~$\Omega$, see Subsection~\ref{180721a}. \smallskip

Suppose to the contrary that the graph $X_\Delta$ is not complete for
any $\Delta\in\pi$. Let~$\Delta$ be a class of $\pi$ that contains a
simplicial vertex of $X$. Then all vertices in $\Delta$ are
simplicial, see~\cite[Lemma 8.1]{forestal_algebras}. It follows that
the graph $X' := X_{ \Omega\setminus\Delta}$ is connected and chordal.
Moreover, the partition $\pi' := \pi_{ \Omega\setminus\Delta}$ is
stable by Lemma~\ref{230621a}(2). Since $|\pi'|<|\pi|$, we conclude by
induction that there is $\Delta'\in\pi'$ such that $X_{\Delta'}'$ is
complete, which is not possible, because $\Delta'\in \pi$.\eprf

\subsection{Estimates depending on the leafage}
The two lemmas in this subsection show bounds that are crucial for
estimating the complexity of the main algorithm.

\lmml{230621y} Let $X$ be a chordal graph, $\Delta$ a subset of its
vertices, $X-\Delta$ is the subgraph of $X$ induced by the complement
of $\Delta$, and \qtnl{230621v}
S=S(X,\Delta)=\{Y\in\con(X-\Delta):\ \ov Y\text{ is not interval}\}.
\eqtn Then $|S|\le\ell(X)-2$.  \elmm

\proof Let $R$ be a tree-representation of $X$ on a tree $T$ such that
$|L(T)|=\ell(X)$. Let $n_3$ be the number of all vertices of $T$ of
degree at least~$3$. Clearly, \qtnl{230621q} n_3\le \ell-2, \eqtn
where $\ell=\ell(X)$. \smallskip


Let $Y\in S$, and let $R(Y)$ be the union of all subtrees $R(\alpha)$,
$\alpha\in\Omega(Y)$. Then $R(Y)$ is a subtree of $T$.  We claim that
$R(Y)$ contains a vertex $t_Y$ of degree at least~$3$. Indeed,
otherwise,~$R(Y)$ is a path in $T$. Moreover, if $\alpha\in\partial
Y$, then either~$R(\alpha)$ is a subpath of $P$, or $R(\alpha)$
contains at least one end of~$P$. This implies that the restriction
of~$R$ to the set $\Omega(Y)\cup\partial Y$ is a tree-representation
of $\ov Y$ on $P$. But then~$\ov Y$ is interval, a
contradiction.\smallskip

To complete the proof, we note that the sets $R(Y)$, $Y\in S$, are   pairwise disjoint. Therefore the vertices $t_Y$ are pairwise distinct. By inequality~\eqref{230621q}, this yields
$$
|S|=|\{t_Y:\ Y\in S\}|\le n_3\le\ell-2,
$$
as required.\eprf\medskip

Let $\pi$ be a vertex coloring of $X$. Given a pair $(\Delta,\Gamma)\in\pi\times \pi$, we define an equivalence relation $e_{\Delta,\Gamma}$ on $\Delta$ by setting
\qtnl{080721a}
(\delta,\delta')\in e_{\Delta,\Gamma}\ \Leftrightarrow\ \delta \text{ and  } \delta' \text{ are twins in } X_{\Delta,\Gamma}.
\eqtn
Note that the equivalence relation $e_{\Gamma,\Delta}$ is defined on $\Gamma$, and coincides with $e_{\Delta,\Gamma}$  only if $\Gamma=\Delta$. The sets of classes of $e_{\Delta,\Gamma}$ and $e_{\Gamma,\Delta}$ are denoted by $\Delta/e_{\Delta,\Gamma}$ and $\Gamma/e_{\Gamma,\Delta}$, respectively.

\lmml{t1}
Let $X$ be a chordal graph, $\pi$ a stable coloring, and $\Delta,\Gamma\in \pi$. Assume that the graph $X_\Delta$ is  complete. Then 
\qtnl{080721v}
|\Delta/e_{\Delta,\Gamma}|\le 2^\ell \quad \text{and} \quad |\Gamma/e_{\Gamma,\Delta}|\leq \ell,
\eqtn
where $\ell = \ell(X)$.
\elmm

\proof Without loss of generality we may assume that
$X=X_{\Delta\cup\Gamma}$ (because
$\ell(X_{\Delta\cup\Gamma})\le\ell(X)$), and the graph
$X_{\Delta,\Gamma}$ is neither complete bipartite, nor empty
(otherwise, $|\Delta/e_{\Delta,\Gamma}|=1$ and
$|\Gamma/e_{\Gamma,\Delta}| = 1$, and both statements are
trivial). Thus,~$X_\Gamma$ is not complete by Lemma~\ref{230621a4}(3)
and $\Delta$ is a maximal clique of $X$; in particular,
$\Delta\ne\Gamma$.\smallskip

Let $R:\Omega\to S(T)$ be a tree-representation of the graph~$X$ on a
tree $T$ with $\ell$ leaves. Without loss of generality, we may assume
that the set $\Omega(T)$ is the minimum possible. Since $\Delta$ is a
clique of $X$, the intersection of the subtrees $R(\delta)$,
$\delta\in\Delta$, contains at least one point~$t$.\smallskip

Let $\gamma\in\Gamma$. Then $t\notin R(\gamma)$ by the maximality of
$\Delta$. Denote by $t_\gamma$ the point of ~$R(\gamma)$, lying at the
minimum distance from $t$ in $T$. Let $P_\gamma$ be the path
connecting~$t$ and $t_\gamma$; note that $P_\gamma$ has at least two
vertices, because $t \neq t_\gamma$. \smallskip

Let us define a partial order on~$\cT = \{t_\gamma : \gamma \in
\Gamma\}$ by setting $t_\gamma\preceq t_{\gamma'}$ if and only if
$t_{\gamma}$ lies in $P_{\gamma'}$ (in particular, either
$t_\gamma=t_{\gamma'}$ or $t_\gamma$ is closer to $t$ than
$t_{\gamma'}$), or equivalently, $P_{\gamma^{}}\subseteq P_{\gamma'}$.

\medskip	
{\noindent\bf Claim.} {If $t_\gamma \preceq t_{\gamma'}$, then $(\gamma,\gamma')\in e_{\Gamma,\Delta}$.}
\medskip

\proof Let $\delta\in \gamma' X\cap \Delta$.  Then the intersection
$R(\delta) \cap R(\gamma')$ is not empty.  Moreover, it contains
$t_{\gamma'}$: for otherwise, because $t\in R(\delta)$, the set
$R(\gamma')$ contains a vertex which is closer to $t$ than
$t_{\gamma'}$. Consequently, $P_{\gamma'} \subseteq R(\delta)$.  Since
$P_{\gamma} \subseteq P_{\gamma'}$ we have
$$
t_\gamma\in P_\gamma\subseteq P_{\gamma'}\subseteq R(\delta).
$$
That is, the intersection $R(\delta)\cap R(\gamma)\ni t_\gamma$ is
 not empty; in particular, $\delta\in\gamma X\cap \Delta$. It follows
  that $ \gamma' X \cap \Delta \subseteq \gamma X \cap \Delta.  $
  Since also $|\gamma' X \cap \Delta|=|\gamma X \cap \Delta|$ by
  Lemma~\ref{230621a}(1), we are done.\eprf\medskip

Let $\cT_{\min} \subseteq \cT$ be the set of all minimal points with
respect to the partial order on~$\cT$. By the claim, for every
$\gamma'\in\cT\setminus \cT_{\min}$ there is $\gamma\in \cT_{\min}$
such that $(\gamma,\gamma')\in e_{\Gamma,\Delta}$.  Thus,
$$
|\Gamma/e_{\Gamma,\Delta}| \leq |\cT_{\min}|.
$$
On the other hand, by the minimality of $T$, every leaf of $T$ belongs to $R(\gamma)$ for some $\gamma\in\Gamma$. Consequently, the path from any leaf of $T$ to $t$ contains at most one point of $\cT_{\min}$. Thus, $|\cT_{\min}|\le \ell$ and so
$$
|\Gamma/e_{\Gamma,\Delta}| \leq |\cT_{\min}|\le \ell,
$$
which proves the second inequality in~\eqref{080721v}.\smallskip

To complete the proof, we observe that if  $\delta\in\Delta$, then the set
$\delta X\cap\Gamma$ is a union of some classes of
$\Gamma/e_{\Gamma,\Delta}$.  Denote this union by $\Gamma_\delta$.
Note that if $\delta,\delta'\in\Delta$, then $\Gamma_{\delta} = \Gamma_{\delta'}$ if and only if $(\delta,\delta')\in
e_{\Delta,\Gamma}$.  Therefore, the number
$|\Delta/e_{\Delta,\Gamma}|$ is at most
$$
|\Delta/e_{\Delta,\Gamma}|\le |2^{\Gamma/e_{\Gamma,\Delta}}|= 2^{|\Gamma/e_{\Gamma,\Delta}|}\le 2^\ell,
$$
which proves the first inequality in~\eqref{080721v}.\eprf

\section{Critical set of a chordal graph}\label{180221z}

Let $X$ be a chordal graph  and $\pi$ a stable  coloring.  Denote by $\Omega^*=\Omega^*(X,\pi)$ the union of all $\Delta\in\pi$ such that 
\qtnl{230621k}
|\con(X_\Delta)|\le \ell(X).
\eqtn
By Lemma~\ref{230621a4}(1), the graph $X_\Delta$ is a disjoint union of cliques; thus the above condition means that the number of them is at most~$\ell(X)$.  By Lemma~\ref{180621c}, the set~$\Omega^*$ is not empty if the graph $X$ is connected. 

\thrml{230621m}

Let $X$ be a chordal graph and $\Omega^*=\Omega^*(X,\pi)$. Then one of
the following statements holds:
\nmrt
\tm{i} for every $Y\in\con(X-\Omega^*)$, the graph $\ov Y_{}$ is interval,
\tm{ii} there is a invariant stable coloring $\pi'>\pi$.
\enmrt
Moreover, in case~(ii), the coloring $\pi'$ can be found in polynomial time
in~$|\Omega|$.
\ethrm

\proof Assume that (i) does not hold. Then the set  the set $S=S(X,\Delta)$ defined
by formula~\eqref{230621v} for $\Delta=\Omega^*$ is not empty. By
Lemma~\ref{230621y}, we have \qtnl{230621h} |S|\le \ell-2, \eqtn where
$\ell=\ell(X)$.  Take an arbitrary $Y\in S$. By
Lemma~\ref{230621a}(2), the coloring $\pi_Y:=\pi_{\Omega(Y)}$ is
stable. By Lemma~\ref{180621c}, there is $\Gamma'\in\pi_Y$ such that
the graph $Y_{\Gamma'}$ is complete. Let~$\Gamma$ be the class
of~$\pi$, containing $\Gamma'$. Then \qtnl{230621h1}
\Gamma\cap\Omega^*=\varnothing, \eqtn because $\Gamma$ intersects
$\Omega\setminus\Omega^*\in \pi^\cup$. Moreover, every automorphism of
$X$ preserves the sets $S$ and $\Gamma$ and hence preserves the set
$$
S'=\{Z\in S:\ Z_{\Gamma\cap \Omega(Z)}\text{ is complete}\}.
$$
Thus the union $\Gamma_0$ of all sets $\Gamma\cap \Omega(Z)$, $Z\in S'$, is a nonempty $\aut(X)$-invariant set contained in~$\Gamma$. Now if $\Gamma_0\ne\Gamma$, then we come to case (ii) with 
$$
\pi'=(\pi\setminus\{\Gamma\})\cup \{\Gamma\setminus\Gamma_0,\Gamma\cap\Gamma_0\}.
$$
To complete the proof, assume that $\Gamma_0=\Gamma$. Then by inequality~\eqref{230621h}, the graph $X_\Gamma$ is the union of at most $|S'|\le|S|\le\ell-2$ cliques. By the definition of $\Omega^*$, this yields $\Gamma\subseteq\Omega^*$, which contradicts  relation~\eqref{230621h1}.\eprf\medskip

We say that $\Omega^*$ is a \emph{critical set} of $X$ (with respect to $\pi$) if statement~(i) of Theorem~\ref{230621m} holds. In the rest of the section we define a hypergraph $\cH^*$ associated with the critical set~$\Omega^*$ and show that the groups $\aut(\cH^*)^{\Omega^*}$ and  $\aut(X)^{\Omega^*}$ are closely related. 

The vertices of $\cH^*$ are set to be the elements of the disjoint  union
$$
V=\bigcup_{ \Delta\in\pi_{\Omega^*}}\bigcup_{\Gamma\in\pi} \Delta/e_{\Delta,\Gamma},
$$
where $e_{\Delta,\Gamma}$ is the equivalence relation on $\Delta$, defined by formula~\eqref{080721a}. Thus any vertex of $\cH^*$ is a class of some $e_{\Delta,\Gamma}$. Taking the disjoint union means, in particular, that if $\Lambda$ is a class of $e_{\Delta,\Gamma^{}}$ and $e_{\Delta,\Gamma'}$, then $V$ contains two vertices corresponding to~$\Lambda$.  The partition 
$$
\ov\pi=\{\Delta/e_{\Delta,\Gamma}:\ \Delta\in\pi_{\Omega^*},\ \Gamma\in\pi\}
$$ 
of the set $V$ is treated as a coloring of~$V$. \smallskip


Let us define the hyperedges of $\cH^*$. First, let  $\alpha\in\Omega^*$. Denote by $\Delta$ the class of~$\pi$, containing $\alpha$. Then $\Delta\in\pi_{\Omega^*}$. Moreover, for every $\Gamma\in \pi$, there is  a unique class $\Lambda_\alpha(\Delta,\Gamma)$ of the equivalence relation~$e_{\Delta,\Gamma}$, containing $\alpha$.  Put
$$
\ov \alpha=\{\Lambda_\alpha(\Delta,\Gamma):\ \Gamma\in\pi\},
$$
in particular, $\ov\alpha\subseteq V$. It is easily seen that
$\ov\alpha=\ov\beta$ if and only if the vertices $\alpha$ and~$\beta$
are twins in $X$, lying in the same class of~$\pi$. Next, let
$\beta\in\Omega^*$ be adjacent to $\alpha$ in~$X$, and $\Gamma$ the
class of $\pi$, containing $\beta$. Then every vertex in
$\Lambda_\alpha(\Delta,\Gamma)$ is adjacent to every vertex of
$\Lambda_\beta(\Gamma,\Delta)$. Put
$$
\ov{\{\alpha,\beta\}}=\{\Lambda_\alpha(\Delta,\Gamma),\Lambda_\beta(\Gamma,\Delta)\},
$$
again $\ov{\{\alpha,\beta\}}\subseteq V$. With this notation, the
hyperedge set of $\cH^*$ is defined as the union:
$$
E^*=\{\ov\alpha:\ \alpha\in \Omega^*\}\,\cup\, \{\ov{\{\alpha,\beta\}}:\ \alpha,\beta\in\Omega^*,\ \beta\in\alpha X\}.
$$

As we are interested in only automorphisms of $E^*$ that stabilize the
two parts $\{\ov\alpha:\ \alpha\in \Omega^*\}$ and
$\{\ov{\{\alpha,\beta\}}:\ \alpha,\beta\in\Omega^*,\ \beta\in\alpha
X\}$, we can color the hyperedges in $E^*$ using two distinct colors
to ensure this. Clearly, the hypergraph $\cH^*=(V,E^*)$ and the
coloring $\ov\pi$ can be constructed in polynomial time in $|\Omega|$.


\thrml{240621a}
Let $X$ be a chordal graph, $\pi$ an invariant stable
vertex coloring of~$X$, $\Omega^*=\Omega^*(X,\pi)$ the critical set,
and $\cH^*=(V,E^*)$ is the above hypergraph with vertex coloring
$\ov\pi$. Then
\nmrt
\tm{i} $\max\{|\Delta|:\ \Delta\in \ov\pi\}\le \ell 2^\ell$, where $\ell=\ell(X)$,
\tm{ii} if $X$ is twinless, then the mapping
$f:\Omega^*\to E^*,\ \alpha\mapsto\ov\alpha$, is an injection,
\tm{iii} if $X$ is twinless and $G=G(\cH^*)$ is the group
induced by the natural action of $\aut(\cH^*)$ on
$\im(f)=\{\ov\alpha\mid \alpha\in\Omega^*\}\subseteq E^*$, then
\qtnl{080721c} \aut(X)^{\Omega^*}\le G^{f^{-1}}\le \aut(X_{\Omega^*}),
\eqtn where $G^{f^{-1}}=fGf^{-1}$.\footnote{Note that the composition
  $fGf^{-1}$ is defined from left to right.}
\enmrt
\ethrm


\proof (i) The color classes of $\ov\pi$ are the sets $\Delta/e_{\Delta,\Gamma}$, where $\Delta\in\pi_{\Omega^*}$ and $\Gamma\in\pi$. By the definition of $\Omega^*$, we have $|\con(X_\Delta)|\le \ell$, and  Lemma~\ref{230621a4}(2) yields 
\qtnl{240621d}
|\con(X_{\Delta\cup\Gamma})|\le\min\{|\con(X_\Delta)|,\,|\con(X_\Gamma)|\}\le \ell.
\eqtn
Further, let $Y\in \con(X_{\Delta\cup\Gamma})$. Then by Lemma~\ref{230621a}(2), the coloring $\pi_Y$ is stable. It has two classes, one inside $\Delta$ and the other one inside $\Gamma$; denote them by $\Delta_Y$  and $\Gamma_Y$, respectively. Note that by  Lemma~\ref{230621a}(2), at least one of the graphs $X_{\Delta_Y}$, $X_{\Gamma_Y}$ is complete. From Lemma~\ref{t1}, we obtain
\qtnl{240621d1}
|\Delta/e_{\Delta,\Gamma}|\le\max\{\ell,2^\ell\}\le 2^\ell.
\eqtn
Since the equivalence relation $e_{\Delta,\Gamma}$ is the union of the equivalence relations $e_{\Delta_Y,\Gamma_Y}$, $Y\in \con(X_{\Delta\cup\Gamma})$, inequalities~\eqref{240621d} and~\eqref{240621d1} imply
$$
|\Delta/e_{\Delta,\Gamma}|=\sum_{Y\in \con(X_{\Delta\cup\Gamma})}|e_{\Delta_Y,\Gamma_Y}|\le \ell\, 2^\ell,
$$
as required.\smallskip

(ii) Assume that $X$ is twinless. Let $\alpha\in\Omega^*$ and let $\Delta\in\pi$ contain $\alpha$. Denote by~$\Lambda_\alpha$ the intersection of all $\Lambda_\alpha(\Delta,\Gamma)$, $\Gamma\in\pi$. Note that every $\beta\in\Lambda_\alpha$ belongs to~$\Delta$. Moreover, 
$$
\alpha X\cap\Gamma=\beta X\cap\Gamma
$$
for all $\Gamma\ne\Delta$, and 
$$
(\alpha X\cap\Delta)\setminus\{\beta\}=(\beta X\cap\Delta)\setminus\{\alpha\}.
$$ 
It follows that $\alpha$ and $\beta$ are twins in $X$. Since $X$ is twinless, we conclude that $\alpha=\beta$. Thus,
$$
\Lambda_\alpha=\{\alpha\}\quad\text{for all }\alpha\in\Omega^*.
$$
Now assume that $f(\alpha)=f(\beta)$ for some $\alpha,\beta\in\Omega^*$. Then $\Lambda_\alpha=\Lambda_\beta$ and the above formula implies $\{\alpha\}=\Lambda_\alpha=\Lambda_\beta=\{\beta\}$. Thus, $\alpha=\beta$ and $f$ is injective.\smallskip

(iii) Assume that $X$ is twinless. By (ii), the mapping $f$ is an injection. Let $g\in\aut(X)$ and $\alpha\in\Omega^*$. Then $\alpha$ lies in some $\Delta\in\pi_{\Omega^*}$. Since $\pi$ is invariant and stable, we have $e_{\Delta^g,\Gamma^g} = e_{\Delta,\Gamma}$ and so
$$
\Lambda_\alpha(\Delta,\Gamma)^{\bar{g}}=\Lambda_{\alpha^g}(\Delta,\Gamma)
$$
for every $\Gamma \in \pi$, where $\bar{g}\in\sym(V)$ is the permutation induced by
$g$. Note that $\bar{g}$ preserves the coloring $\ov \pi$. Moreover,
$$
(\ov\alpha)^{\bar{g}}=\{\Lambda_\alpha(\Delta,\Gamma)^{\bar{g}}:\ \Gamma\in\pi\}=\{\Lambda_{\alpha^g}(\Delta,\Gamma):\ \Gamma\in\pi\}=\ov{\alpha^g}
$$
and
\qtnl{250621a}
\ov{\{\alpha,\beta\}}^{\bar{g}}=\{\Lambda_\alpha(\Delta,\Gamma)^{\bar{g}},\Lambda_\beta(\Gamma,\Delta)^{\bar{g}}\}=
\{\Lambda_{\alpha^g}(\Delta,\Gamma),\Lambda_{\beta^g}(\Gamma,\Delta)\}=
\ov{\{\alpha^g,\beta^g\}}.
\eqtn
Consequently, $\bar{g}\in\aut(\cH^*)$. Since ${\ov
  \alpha}^{f^{-1}gf}=(\alpha^g)^f = \ov {\alpha^g}= {\ov \alpha}^{\bar{g}}$,
it follows that $f^{-1}gf\in G$, which proves the left-hand side inclusion
in~\eqref{080721c}.

Let $\alpha,\beta\in\Omega^*$.  Denote by $\Delta$ and $\Gamma$ the classes of $\pi$, containing $\alpha$ and $\beta$, respectively. Then $\alpha$ and $\beta$ are adjacent in $X$ if and only if every vertex in $\Lambda_\alpha(\Delta,\Gamma)$ is adjacent to every vertex of $\Lambda_\beta(\Gamma,\Delta)$, or equivalently, $\ov{\{\alpha,\beta\}}\in E^*$. Thus, the right-hand side inclusion in~\eqref{080721c} follows from~\eqref{250621a}.\eprf

\section{The hypergraph associated with complement of the critical set}\label{250621q}

The goal of this section is to provide some tools related to the
critical set that will help design the algorithm for computing the
automorphism group of a chordal graph in $\fK_\ell$.

Suppose $X$ is a chordal graph on $\Omega$ and $\pi$ an invariant
stable coloring of~$X$. Further, let $\Omega^*$ denote the critical
set of~$X$ with respect to~$\pi$. Let $G^\diamond=G^\diamond(X)$
denote the kernel of the restriction homomorphism $\aut(X)\to
\aut(X)^{\Omega^*}$. We claim that a generating set for $G^\diamond$
can be efficiently computed.

\thrml{250627s}

A generating set for the kernel $G^\diamond\le \sym(\Omega)$ of the
restriction homomorphism from $\aut(X)$ to $\aut(X)^{\Omega^*}$ can be
found in polynomial time in~$|\Omega|$.

\ethrm

\proof Without loss of generality, we may assume that the set
$\Omega^\diamond=\Omega\setminus\Omega^*$ is not empty. Let us define
a vertex coloring $\pi^\diamond$ of the graph
$X^\diamond=X_{\Omega^\diamond}$, such that
$\pi^\diamond(\alpha)=\pi^\diamond(\beta)$ if and only if
$\pi(\alpha)=\pi(\beta)$ and $\alpha X\cap \Omega^* = \beta X\cap
\Omega^*$. It is not hard to see that
$$
(G^\diamond)^{\Omega^\diamond} = \aut(X^\diamond,\pi^\diamond).
$$
Since also the graph $X^\diamond$ is interval (see the definition of
the critical set), a generating set of
$(G^\diamond)^{\Omega^\diamond}$ can be found by the algorithm
in~\cite[Theorem 3.4]{colbourn_booth}, which constructs a generating
set of the automorphism group of a vertex colored interval graph
efficiently. Since $(G^\diamond)^{\Omega^*} =\{\id_{\Omega^*}\}$, the theorem
is proved.\eprf\medskip

In what follows, $X$ is a chordal graph, $\pi$ a stable coloring of
$X$, $\Omega^*$ the critical set of~$X$ with respect to~$\pi$, and
$\Omega^\diamond = \Omega\setminus \Omega^*$. Recall that by the
definition of critical set, every graph $\ov Y$, $Y\in
\con(X_{\Omega^\diamond})$, is interval and
$$
\partial Y=\Omega(\ov Y)\cap\Omega^*.
$$

\lmml{300621a} For every $Y\in \con(X_{\Omega^\diamond})$, there is a
colored hypergraph $H=H_Y$ whose vertex set is $\partial Y$ colored by
$\pi_{\partial Y}$, and such that \qtnl{280621a}
\iso(H_{Y^{}},H_{Y'})=\iso(\ov Y,\ov {Y'})^{\partial Y}, \quad Y'\in
\con(X_{\Omega^\diamond}).  \eqtn

Moreover, in time polynomial in $|\ov Y|$ one can
\nmrt
\tm{a} construct the hypergraph $H_Y$,
\tm{b} given $\ov g\in\iso(H_{Y^{}},H_{Y'})$, find
$g\in \iso(\ov Y,\ov {Y'})$
such that $g^{\partial Y}=\ov g$.
\enmrt
\elmm


\proof We make use of the results of \cite{forestal_algebras}. Namely,
let $Z$ be an interval graph and~$\pi_Z$ a stable coloring of $Z$.
From \cite[Theorem~6.10 and Proposition~6.4]{forestal_algebras}, it
follows that there exists a canonical rooted tree $T=T(Z)$ and a
stable coloring $\pi_T$ of $T$ such that \qtnl{300621w}
L(T)=\Omega(Z)\qaq \pi_Z=(\pi_T)_{L(T)}.  \eqtn The term ``canonical''
means that for every interval vertex colored graph $Z'$, the
isomorphisms between $Z$ and $Z'$ are related with the isomorphisms
between $T$ and $T'=T(Z')$ as follows: \qtnl{300621w1}
\iso(T,T')^{L(T)}=\iso(Z,Z').  \eqtn Moreover, the proof of
\cite[Proposition~6.4]{forestal_algebras} shows that the sizes of $T$
and $\pi_T$ are polynomials in $|\Omega(Z)|$, and $T$ and $\pi_T$ can
be constructed in polynomial time.\smallskip

Now let $Y\in \con(X_{\Omega^\diamond})$. Since the graph $\ov Y$ is interval, one can define the rooted tree $T=T(\ov Y)$ as above. Next, for each vertex $x$ of $T$, we introduce the 
following notation:
\nmrt
\item[$\bullet$] $L(x)$ is the set of all descendants of $x$ in~$T$, lying in $L(T)$,  
\item[$\bullet$] $T_x$ is the subtree of $T$ rooted at~$x$ and such that $L(T_x)=L(x)\setminus\partial Y$, 
\item[$\bullet$] $F(T_x)$ is a string encoding the isomorphism type of
  the rooted tree~$T_x$.  \enmrt Now for each $x$ with
  $L(x)\cap\partial Y\ne\varnothing$ and $L(x)\setminus\partial
  Y\ne\varnothing$, we delete from $T$ all the vertices of $T_x$,
  except for~$x$, and define the new color of $x$ to be equal to
  $(\pi_T(x),F(T_x))$. Denote the resulting tree and its vertex
  coloring by $T_1=T_1(Y)$ and $\pi_1=\pi_1(Y)$, respectively.  Then
  \qtnl{010721a1} L(T_1)=\partial Y.  \eqtn It is not hard to see that
  $T_1$ and $\pi_1$ can efficiently be constructed, and $T$ and
  $\pi_T$ are uniquely recovered from $T_1$ and $\pi_1$. In
  particular, \qtnl{010721a} \iso(T^{}_1,T'_1)^{\partial Y}=\iso(\ov
  Y,\ov Y'), \eqtn where $Y'\in\con(X-\Omega^*)$ and $T'_1=T_1(Y')$,
  respectively, cfg., \eqref{300621w} and \eqref{300621w1}.\smallskip

At this point we can define the required hypergraph $H_Y=(\partial Y,E_Y)$, where
$$
E_Y=\{L(x):\ x\in\Omega(T_1(Y))\}.
$$
Note that $L(x)=L(y)$ if and only if $x=y$ or one of $x,y$ is  the descendants of the other in~$T_1$, and if, say $y$ is the descendant of $x$, then for each vertex $z\ne y$ of the path $P_{yx}$ connecting $y$ with $x$, we have $L(x)=L(z)$; moreover, in the latter case,~$z$ has a unique child in $T_1$. Thus for each $e\in E_Y$ there exist uniquely determined vertex $x_e$ and its descendant $y_e$ such that 
$$
L(z)=e\ \Leftrightarrow\ z\in\Omega(P_{y_ex_e})\text{ and } z\ne y_e \text{ if }x_e\ne y_e.
$$
In particular, if $x_e=y_e:=x$, then $L(x)=L(z)$ if and only if $x=z$.\smallskip

To define the color of the hyperedge $e\in E_Y$, let
$\Omega(P_{y_ex_e})=\{y_0, y_1,\ldots y_k\}$, where $k\ge 0$ is the
length of $P_{y_ex_e}$, $y_0=y_e$, $y_k=x_e$, and $y_i$ is the child
of $y_{i+1}$, $i=0,\ldots,k-1$. Then the color of $e$ is set to be the
tuple
$$
(\pi_1(y_1),\ldots,\pi_1(y_k)).
$$ 
Again, it is clear that the hypergraph $H_Y$ and its coloring can be
constructed efficiently and that they determine the colored tree $T_1$
in a unique way. Thus the statement of the lemma is a consequence of
formulas~\eqref{010721a1} and~\eqref{010721a}.\eprf\medskip

Let us define a colored \order{2} hypergraph $\cH^\diamond$ with vertex set $\Omega^*$ and hyperedge set $\cE_1\cup\cE_2$, where
$$
\cE_1=\bigcup_{Y\in \con(X_{\Omega^\diamond})}E(H_Y)\qaq
\cE_2=\{E(H_Y):\ Y\in \con(X_{\Omega^\diamond}\}.
$$

The vertex coloring of $\cH^\diamond$ is set to be
$\pi_{\Omega^*}$. Note that the union in the definition of~$\cE_1$ is
not disjoint; the color $\pi^\diamond(e)$ of a hyperedge $e\in \cE_1$ is
defined to be the multiset of the  colors of~$e$ in
$\cH_Y$, where $Y$ runs over all graphs $Y\in
\con(X_{\Omega^\diamond})$ such that $e\in E(H_Y)$.\smallskip

To define a coloring of $\cE_2$, denote by $\sim$ the equivalence
relation on $\con(X_{\Omega^\diamond})$ by setting
$$
Y\sim Y'\quad\Leftrightarrow\quad H_{Y^{}}=H_{Y'}.
$$ 
Condition~\eqref{280621a} implies that $Y\sim Y'$ if and only if there exists an isomorphism $g\in \iso(\ov Y,\ov Y\phmb{'})$ such that the bijection $g^{\partial Y}$ is identical. The color $\pi^\diamond(e)$ of the hyperedge $e\in \cE_2$ is defined to be so that
if $e=\{E(H_{Y^{}})\}$ and $e'=\{E(H_{Y'})\}$, then
\qtnl{300621c}
\pi^\diamond(e)=\pi^\diamond(e')\quad\Leftrightarrow\quad \iso(\ov Y,\ov Y')\ne\varnothing\qaq n_{Y^{}}=n_{Y'},
\eqtn
where $n_{Y^{}}$ and $n_{Y'}$ are the cardinalities of the  classes of the equivalence relation~$\sim$, containing $Y$ and $Y'$, respectively.

\rmrkl{020721y}
Let $e\in\cE_2$ and $Y\in\con(X-\Omega^*)$ be such that $e=E(H_Y)$. In general, the coloring $\pi_e$ of the hyperedges of $\cE_1$, contained in $e$, is different from the coloring $\pi_Y$ of the corresponding hyperedges of $H_Y$. However, $\pi_e\ge \pi_Y$ and $\pi_Y$ is uniquely determined by~$\pi_e$. 
\ermrk

\lmml{280621d}
Let $X'$ be a colored graph obtained from $X$ by deleting all edges of the induced subgraph $X_{\Omega^*}$. Then  
$$
\aut(\cH^\diamond)=\aut(X')^{\Omega^*}.
$$
Moreover, given $\ov g\in \aut(\cH^\diamond)$ one can construct $g\in \aut(X')$ such that $g^{\Omega^*}=\ov g$ in polynomial time in $|\Omega|$.
\elmm
\proof Let $g\in\aut(X')$. Since the set $\Omega^*$ is  $\aut(X')$-invariant, the permutation~$\ov g=g^{\Omega^*}$ preserves the coloring $\pi_{\Omega^*}$. Moreover, $g$ induces a permutation 
\qtnl{180721q}
Y\mapsto Y',\quad Y\in\con(X-\Omega^*),
\eqtn 
such that $(\partial Y)^g=\partial Y'$ for all $Y$, and the  isomorphisms
$$
g_Y\in\iso(\ov Y,\ov {Y'}),\quad Y\in\con(X-\Omega^*).
$$
By formula~\eqref{280621a}, we have  $(g_Y)^{\partial Y}=g^{\partial Y}\in \iso(H_{Y^{}},H_{Y'})$. Now, let  $e\in\cE_1$. Then $e\in E(H_Y)$ for some $Y\in\con(X-\Omega^*)$. It follows that
$$
e^g=e^{g_Y}\in E(H_{Y'})\quad\text{for all}\ \,e\in E(H_Y).
$$
Consequently, the permutation $\ov g$ preserves the hyperedges of $\cE_1$. Because the isomorphism  $g_Y$ is color preserving, $\ov g$ preserves also the colors of them. Finally, the automorphism $g\in\aut(X')$ preserves the relations on the right-hand side of formula~\eqref{300621c} and hence the permutation \eqref{180721q} leaves the equivalence relation~$\sim$ fixed. Since $g$ induces the same permutation, we conclude that~$\ov g$ preserves the colors of the hyperedges of~$\cE_2$. Thus, $\ov g\in\aut(\cH^\diamond)$.\smallskip

Conversely, let $\ov g\in\aut\cH^\diamond$. Formula~\eqref{300621c} implies that $\ov g$ induces a cardinality preserving permutation of the classes of the equivalence relation~$\sim$. Consequently, there is a permutation~\eqref{180721q} such that $\iso(H_{Y^{}},H_{Y'})\ne\varnothing$; although such a permutation is not necessarily unique, one can efficiently find at least one such permutation.\smallskip

Recall that $\cE_1^{\ov g}=\cE_1$. Moreover, the hyperedges  from  $E(H_Y)\in\cE_2$ go to the edges from $E(H_{Y'})\in\cE_2$. Therefore (see Remark~\ref{020721y}), 
\qtnl{300621h}
\ov g_Y:={\ov g}\phmb{\partial Y}\in\iso(H_{Y^{}},H_{Y'}).
\eqtn
By formula~\eqref{280621a}, there exists a bijection $g_Y\in\iso(\ov Y,\ov{Y'})$ such that
\qtnl{300621h1}
g_Y^{\partial Y}=\ov g_Y,
\eqtn
and this bijection can efficiently be found
(Lemma~\ref{300621a}(b)). Now we define a permutation
$g\in\sym(\Omega)$ by setting $\alpha^g=\alpha^{g_Y}$, where $Y$ is an
arbitrary element of~$\con(X-\Omega^*)$, for which
$\alpha\in\Omega(\ov Y)$. The permutation $g$ is well defined, because
by \eqref{300621h} and \eqref{300621h1},
$$
\alpha^{g_Y}=\alpha^{\ov g_Y}=\alpha^{\ov g}=\alpha^{\ov g_Z}=\alpha^{g_Z}
$$
for all $Z\in\con(X-\Omega^*)$ and all $\alpha\in\partial Y\cap\partial Z$. It remains to note that $g\in\aut(X')$, because $g$ moves edges of each $\ov Y$ to $\ov{Y'}$, and $E(X')$ is the union of the sets~$E(\ov Y)$.\eprf\medskip

The following theorem is the main result of the section, which together with Theorem~\ref{250627s} essentially provides a polynomial-time reduction of finding the group $\aut(X)$ to finding the groups $\aut(\cH^*)$ and $\aut(\cH^\diamond)$.

\thrml{250627r}
In the conditions and notation of Theorem~\ref{240621a}, set
$G^*=G(\cH^*)^{f^{-1}}$.  Then
$$
\aut(X)^{\Omega^*}= \aut(\cH^\diamond)\cap G^*. 
$$
Moreover, every permutation $\ov g\in \aut(\cH^\diamond) \cap G^*$
can be lifted in polynomial time to an automorphism $g\in\aut(X)$ such
that $g^{\Omega^*}=g$.
\ethrm

\proof

By Theorem~\ref{240621a}, we have $\aut(X)^{\Omega^*} \leq
G^*$. Furthermore, $\aut(X)\le\aut(X')$, where $X'$
is the graph from Lemma~\ref{280621d}. By that lemma, this implies
that
$\aut(X)^{\Omega^*}\le\aut(X')^{\Omega^*}=\aut(\cH^\diamond)$. Thus,
$$
\aut(X)^{\Omega^*}\le \aut(\cH^\diamond) \cap G^*.
$$

Conversely, let $\ov g\in \aut(\cH^\diamond)\cap G^*$. By
Lemma~\ref{280621d}, one can efficiently find $g\in\aut(X')$ such that
$g^{\Omega^*}=\ov g$. Now, by Theorem~\ref{240621a} the
permutation~$g$ preserves the edges of $X$ contained in
$E(X_{\Omega^*})$. The other edges of $X$ are exactly those in $E(X')$
and $g$ preserves them by Lemma~\ref{280621d}. Thus,
$$
E(X)^g=(E(X_{\Omega^*})\,\cup\,E(X'))^g=E(X_{\Omega^*})^g\,\cup\,E(X')^g=
E(X_{\Omega^*})\,\cup\,E(X')=E(X),
$$
i.e., $g\in \aut(X)$, as required.\eprf

\section{Order-$k$ hypergraph isomorphism: bounded color classes}\label{250621y}

The goal of this section is to design an FPT algorithm for testing
isomorphism of colored $k$-hypergraphs in which the sizes of vertex
color classes are bounded by a fixed parameter; no assumption is made
on the hyperedge color class sizes. The algorithm we present is a generalization of the one for usual hypergraphs~\cite{colored_hypergraph_fpt}.

\thrml{100721a}
Let $k\ge 1$. Given two colored \order{k} hypergraphs
$H$ and $H'$, the isomorphism coset $\iso(H,H')$ can be computed in
time $(b!\,s)^{O(k)}$, where $b$ is the maximal size of a vertex color
class of $H$ and $s$ is the size of $H$. In particular, the group
$\aut(H)$ can be found within the same time.
\ethrm

The proof of Theorem~\ref{100721a} is given at the end of the
section. We start with some notation and definitions; most of them go
back to those in~~\cite{colored_hypergraph_fpt}. In what follows, we
fix a finite set $V$ and the decomposition of $V$ into the disjoint
union of its color classes,
\qtnl{110721q}
V = C_1\sqcup C_2\sqcup\cdots \sqcup C_m,
\eqtn
where $m\ge 1$ and $|C_i|\le b$ for each $i$. For every higher order
hyperedge $e$, we consider its projections to unions of the color
classes,

$$
e^{(i)}=e^{C_1\cup C_2\cup\cdots \cup C_i},\quad 0\le i\le m,
$$
see Subsection~\ref{hohg-defn}. Obviously, $e^{(0)}=\varnothing$ and
$e^{(m)}=e$.\smallskip

{\bf $i$-equivalence.} Let $i\in\{0,\ldots,m\}$. Two \order{k}
hyperedges $e$ and $e'$ are said to be \emph{$i$-equivalent} if the
multisets $e^{(i)}$ and ${e'}\phmb{(i)}$ are equal. The following
statement is straightforward.

\prpstnl{180721z}
\phantom{x}
\nmrt 
\tm{1} any two high order hyperedges are $0$-equivalent,
\tm{2} for $i\ge 1$, any two $i$-equivalent  high order hyperedges are $(i-1)$-equivalent,
\tm{3} two high order hyperedges are $m$-equivalent if and only if they are equal.
\enmrt
\eprpstn

{\bf $i$-blocks.} Let $H=(V,E)$ be an \order{k} hypergraph. For every
$i\in\{0,\ldots,m\}$, the $i$-equivalence partitions the set $E$ into
equivalence classes called \emph{$i$-blocks}; the set of all of them
is denoted by $\hat{E}_i$. From Proposition~\ref{180721z}, it
follows that
\qtnl{180721y}
\wh E_0=\{E\}\qaq \wh E_m=E.
\eqtn

{\bf Hypergraphs $A[i]$ associated with $i$-blocks.} Each $i$-block $A\in\wh E_i$
defines an \order{k} hypergraph $(V,A)$, which is just $H$ if
$i=0$, and is essentially the \order{(k-1)} hypergraph $(V,e)$ if
$i=m$ and $A=\{e\}$ for some~$ e\in E$.  Denote by $A[i]$ the 
\order{k} hypergraph on the set
$$
V_i= C_i\sqcup C_{i+1}\sqcup \cdots \sqcup C_m,
$$
obtained from the projection $A^{V_i}$ of $A$ to $V_i$ by replacing each multiset
$e^{V_i}$, $e\in A$ with the corresponding set (without repetitions).
Then $A[0]=H$.\smallskip

{\bf Coloring of $A[i]$.} Assume that the hypergraph $H$ is colored.
The vertex coloring of the hypergraph $A[i]$ is defined in a natural
way, whereas the color of the hyperedge corresponding to $e^{V_i}$ is
defined as a multiset
 $$
\{\{\pi(\wt e):\ \wt e\phmb{\,V_i}=e^{V_i},\ \wt e\in A\}\},
$$ 
where $\pi$ is the coloring of $E(H)$.\smallskip


{\bf Proof of Theorem~\ref{100721a}.} Let $H=(V,E)$ and $H'=(V',E')$ be colored \order{k}
hypergraphs. Without loss of generality we may assume that there is a
decomposition of $V'$ similar to~\eqref{110721q} with the same $m$ and
$b$. Our aim is to design an algorithm of running time
$x(k,s,b)=(b!\,s)^{O(k)}$ for computing the coset $\iso(H,H')$.\smallskip

 Inductively, assume that $k\ge 2$ and we have such an algorithm for order-$(k-1)$
 hypergraphs of  running time $x(k-1,s,b)$. As the base case for the
 induction, by \cite[Corollary~9]{colored_hypergraph_fpt}, we already
 have
\qtnl{180721f}
x(1,s,b)=2^{O(b)}\poly(s).
\eqtn
The algorithm for \order{k} hypergraphs will invoke as subroutine the
algorithm for \order{(k-1)} hypergraphs. Put
$$
C(k,i;H,H')=\{\iso(A[i],A'[i]):\ A\in \wh E_i,\ A'\in \wh E'_i\},\quad 0\le i\le m.
$$

The algorithm below computes the collections $C(k,i;H,H')$ for
decreasing values of $i$ from $m$ down to $0$. Specifically, for each
$i$, it first computes the set $C(k,i+1;H,H')$ and  uses it  for computing the set $C(k,i;H,H')$. Since $A[0]=H$ and
$A'[0]=H'$, notice that we will finally have computed
$\iso(H,H')=C(k,0;H,H')$ as required.\medskip

\centerline{\bf Algorithm for computing $C(k,0;H,H')$}\medskip

\noindent{\bf Input:} colored \order{k} hypergraphs $H=(V,E)$ and $H'=(V',E')$,
$k>1$.\smallskip

\noindent{\bf Output:} the table of all $C(k,i;H,H'), 0\le i\le m$.\medskip

\noindent{\bf For} $i:=m$ {\bf down to} $1$ {\bf do}\smallskip

for all  $A\in \wh E_i$ and $A'\in \wh E'_i$ add to $C(k,i;H,H')$ the coset $\iso(A[i],A'[i])$ computed below. \smallskip

\noindent{\bf Step 0.} \textbf{If} $i=m$\smallskip

\textbf{then} $A[i]$ and $A'[i]$ are \order{k} hypergraphs on the sets
$C_m$ and~$C'_m$ of cardinality at most~$b$. In this case, $\iso(A[i],A'[i])$ can be
computed in time $O(b!s)$ by inspecting all bijections from $V_i$ to $V'_i$.\smallskip

\textbf{else}\smallskip

\noindent{\bf Step 1.} Construct the $(k-1)$-skeleton hypergraphs $Y=A[i]^{(k-1)}$ and $Y'=A'[i]^{(k-1)}$ (see Section~\ref{hohg-defn}). \smallskip

\noindent{\bf Step 2.} Compute $K\tau:=\iso(Y,Y')=C(k-1,0;Y,Y')$ by using the algorithm for \order{(k-1)} hypergraphs as subroutine.\smallskip

\noindent{\bf Step 3.} Computation of $\iso(A[i],A'[i])$:\smallskip


{\bf Step 3.1.} Let $A_1,A_2,\ldots,A_\ell$ and $A'_1,A'_2,\ldots,A'_{\ell'}$ be the $(i+1)$-blocks 
contained in $A$ and $A'$, respectively; if $\ell\ne\ell'$, then  set $\iso(A[i],A'[i])=\varnothing$.\smallskip

{\bf Step 3.2.} Find the set $P\le\sym(\ell)$ of all permutations  induced by $K\tau$ as the bijections from
$C_{i+1}$ to $C'_{i+1}$ which maps the set $\{A_1,A_2,\ldots,A_\ell\}$ to $\{A'_1,A'_2,\ldots,A'_\ell\}$; note that $|P|\le b!$. \smallskip

{\bf Step 3.3.} Using the algorithm in
\cite[Theorem~5]{colored_hypergraph_fpt}, compute the coset
\qtnl{110721w} 
\iso(A[i],A'[i]) = \bigcup_{\pi\in P}\bigcap_{j=1}^\ell 
\iso(A_j[i+1],A'_{\pi(j)}[i+1]) ,
\eqtn 
where the cosets on the right-hand side  are available from the set $C(k,i+1;H,H')$ found earlier.\smallskip
 
{\bf end-for}\eprf
\medskip

\noindent{\bf Correctness and Analysis.}~ By induction, it suffices to
see how Step 3 computes $\iso(A[i],A'[i])$. Notice that the union on
the right-hand side of~\eqref{110721w} with $P$ replaced by the set of
\emph{all} bijections from~$C_{i+1}$ to $C'_{i+1}$ gives the coset
$H\nu$ of all isomorphisms from $A[i]$ to~$A'[i]$ \emph{projected} to
$V^{}_{i+1}$ and $V'_{i+1}$. Since $A$ and $A'$ are $i$-blocks, they
are single \order{k} hyperedges on color class $C_i$ and $C'_i$,
respectively. In view of formula~\eqref{110721a}, the coset
$K\tau=\iso(Y,Y')$ restricted to $C_i$ and $C'_i$ precisely includes
all the isomorphisms from $A[i]$ to $A'[i]$ restricted to $C_i$ and
$C'_i$. Hence, $K\tau\cap H\nu$ is precisely $\iso(A[i],A'[i])$ which
is computed at Steps~3.3.\smallskip

We analyze the running time $x(k,s,b)$ for the computation of the
set $C(k,0;H,H')$. The outer for-loop executes $m$ times and the inner
for-loop executes at most $|E|^2$ times (for each pair $A$, $A'$ of $i$-blocks).\smallskip

We now bound the time required for computing each $C(k,i;H,H')$. By
induction, each iteration of Steps 0-2 require time 
$$
O(|E|^2\cdot b!)+ x(k-1,s,b)+\poly(s).
$$ 
The number $\ell$ in Step~3.1 is at most
$|E|$. Therefore the cost of Steps~3.1-3.2 is at most
$|E|\,|P|\,\poly(s)\le b!\poly(s)$. Finally, in Step~3.3, we compute
at most $b!$ intersections of $\ell$ cosets available in the
already computed set $C(k,i+1,H,H')$.  Since the intersection of two such
cosets by the algorithm from \cite[Theorem~5]{colored_hypergraph_fpt}
requires $2^{O(b)}\cdot\poly(s)$ time, the overall cost of Step~3 is
at most $O(b!)\poly(s)$. Putting it together, the time spent in
computing $C(k,i;H,H')$, given the pre-computed table entries for
$C(k,i+1,H,H')$, is bounded by $x(k-1,s,b)\cdot O(b!)\poly(s)$.
It follows that the overall time for computing $C(k;H,H')$ is
bounded by $m\cdot |E|^2\cdot x(k-1,s,b)\cdot O(b!)\poly(s)$.
Thus, we have
\[
x(k,s,b)\le m\cdot |E|^2\cdot x(k-1,s,b)\cdot O(b!)\poly(s)\le x(k-1,s,b)\cdot (b!\cdot s)^c,
\]
for a suitable constant $c>0$. By induction hypothesis $x(k-1,s,b)\le
(b!\cdot s)^{c\cdot(k-1)}$. Hence, we obtain an overall upper bound of
$(b!\cdot s)^{c\cdot k}$ for the running time of the algorithm for
\order{k} hypergraphs.
\eprf

\section{Main algorithm and the proof of Theorem~\ref{180221f}}\label{010721z}

Based on the results obtained in the previous sections, we present an algorithm that constructs the automorphism group of a chordal twinless graph.\medskip

\centerline{\bf Main Algorithm}\medskip

{\bf Input:} a chordal twinless graph $X$ and vertex coloring $\pi$ of $X$.

\noindent{\bf Output:} the group $\aut(X,\pi)$.\medskip

\noindent{\bf Step 1.} Construct $\pi=\WL(X,\pi)$ and $\Omega^*=\Omega^*(X,\pi)$.\smallskip

\noindent{\bf Step 2.}  While the set $\Omega^*$ is not critical with respect to~$\pi$, find $\pi:=\WL(X,\pi')$ and set $\Omega^*:=\Omega^*(X,\pi)$, where $\pi'$ is the coloring from Theorem~\ref{230621m}(ii).\smallskip

\noindent{\bf Step 3.} If $\Omega^*=\varnothing$, then $X$ is interval and we output the group $\aut(X,\pi)$ found by the algorithm from \cite[Theorem 5]{linear_interval_isomorphism}.\smallskip

\noindent{\bf Step 4.} Construct the mapping~$f$ and colored
hypergraph $\cH^*$ on $(\Omega^*)^f$, defined in
Section~\ref{180221z}, and the colored hypergraph $H^\diamond$ on
$\Omega^*$, defined in Section~\ref{250621q}.\smallskip

\noindent{\bf Step 5.} Using the algorithm from Theorem~\ref{100721a}, find a generating set $\ov S$ of the automorphism group of the colored \order{3} hypergraph $\cH^*\uparrow(\cH^\diamond)^f$.\smallskip 

\noindent{\bf Step 6.} For each $\ov g\in \ov S$ find a lifting $g\in \aut(X,\pi)$ of  $f\ov gf^{-1}\in\sym(\Omega^*)$ by  the algorithm from Theorem~\ref{250627r}; let $S$ be the set of all these automorphisms~$g$'s.\smallskip

\noindent{\bf Step 7.}  Output the group $\aut(X,\pi)=\grp{G^\diamond,S}$, where $G^\diamond$ is the group defined in Theorem~\ref{250627s}.\eprf\medskip

\thrml{010721e}
The Main Algorithm correctly finds the group $\aut(X,\pi)$ in time
$t(\ell)\cdot n^{O(1)}$, where $n=|\Omega(X)|$, $t$ is a
function independent of~$n$, and $\ell=\ell(X)$.

\ethrm \proof Note that the number of iterations of the loop at Step~2
is at most $n$, because $|\pi|\le n$ and $|\pi'|>|\pi|$. Next, the
running time at each other step, except for Step~5, is bounded by a
polynomial in $n$, see the time bounds in the used statements. On the
other hand, at Step~5, the cardinality of each vertex color class of
the \order{3} hypergraph $\cH^*\uparrow(\cH^\diamond)^f$ is at most
$\ell 2^\ell$ (Theorem~\ref{240621a}(i)). By Theorem~\ref{100721a} for
$b=\ell 2^\ell$ and $k=3$, the running time of the Main Algorithm is at most
$t(\ell)\cdot n^{O(1)}$ with $t(\ell)=((\ell 2^\ell)!)^{O(1)}$.\smallskip

To prove the correctness of the algorithm, we exploit the natural
restriction homomorphism
$$
\varphi:\aut(X)\to \sym(\Omega^*),\ g\mapsto g^{\Omega^*}.
$$
Given  a generating  set $S'$  of the group $\im(\varphi)$, we have $\aut(X)=\grp{\ker(\varphi),S}$, where $S\subseteq\aut(X)$ is a set of cardinality $|\ov S|$ such that $S'=\{\varphi(g):\ g\in S\}$. \smallskip

According to Step~7, $\ker(\varphi)=G^\diamond$. Thus, it suffices to verify that as the set $S'$ one can take the set $\{f\ov gf^{-1}:\ \ov g\in \ov S\}$, where $f$ is the bijection found at Step~$4$ and~$\ov S$ is the generating set of the group $\aut(\cH^*\uparrow(\cH^\diamond)^f)$, found at Step~5. By Theorem~\ref{250627r}, we need to check that
\qtnl{010721f}
\aut(\cH^*\uparrow(\cH^\diamond)^f)^{f^{-1}}= G^*\cap \aut(\cH^\diamond).
\eqtn
Notice that  
\begin{align*}
h\in \aut(\cH^*\uparrow(\cH^\diamond)^f)&\ \Leftrightarrow\ 
h\in \aut(\cH^*)\qaq (E(\cH^\diamond)^f)^h=E(\cH^\diamond)^f\\
&\ \Leftrightarrow\  fhf^{-1}\in G^*\qaq fhf^{-1}\in \aut(\cH^\diamond)\\ &\ \Leftrightarrow\ 
fhf^{-1}\in G^*\cap  \aut(\cH^\diamond),
\end{align*}
which proves equality~\eqref{010721f}.\eprf\medskip

{\bf Proof of Theorem~\ref{180221f}.} Denote by $e_X$ the equivalence
relation on $\Omega=\Omega(X)$ such that $(\alpha,\beta)\in e_X$ if
and only if the vertices $\alpha$ and $\beta$ are twins in $X$. Since
$e_X$ is $\aut(X)$-invariant, there is a natural homomorphism
$$
\varphi:\aut(X)\to\sym(\Omega/e_X).
$$
To find the group $\aut(X)$, it suffices to construct generating sets
of the groups $\ker(\varphi)$ and $\im(\varphi)$, and then to lift
every generator of the latter to an automorphism of~$X$. \smallskip


First, we note that every class of the equivalence relation $e_X$
consists of twins of~$X$. Consequently,
$$
\ker(\varphi)=\prod_{\Delta\in\Omega/e_X}\sym(\Delta),
$$
and this group can efficiently be found.\smallskip

Now let $X'$ be the graph with vertex set $\Omega/e$, in which the
classes $\Delta$ and $\Gamma$ are adjacent if and only if some (and
hence each) vertex in $\Delta$ is adjacent to some (and hence each)
vertex of $\Gamma$. Note that $X'$ is isomorphic to an induced
subgraph of $X$, and hence belongs to the class $\fK_\ell$. Let $\pi'$
be the vertex coloring of $X'$ such that $\pi'(\Delta)=\pi'(\Gamma)$
if and only if $X_\Delta$ and $X_\Gamma$ are isomorphic, which is easy
to check because each of $X_\Delta$ and $X_\Gamma$ is either empty or
complete. Then
$$
\im(\varphi)=\aut(X',\pi'),
$$
and this group can efficiently be found in time  $t(\ell)\cdot
  n^{O(1)}$ by Theorem~\ref{010721e}.\smallskip

To complete the proof, we need to show that given $g'\in
\aut(X',\pi')$, one can efficiently find $g\in\aut(X)$ such that
$\varphi(g)=g'$. To this end, choose an arbitrary bijection
$g_\Delta:\Delta\to\Delta^{\ov g}$; recall that
$\pi'(\Delta)=\pi'(\Delta^{\ov g})$ and so $|\Delta|=|\Delta^{\ov
  g}|$. Then the mapping $g$ taking a vertex $\alpha\in \Omega$ to the
vertex $\alpha^{g_\Delta}$, where $\Delta$ is the class of $e_X$,
containing~$\alpha$ is a permutation of $\Omega$. Moreover, from the
definition of $e_X$, it follows that $g\in\aut(X)$. It remains to note
that $g$ can efficiently be constructed.\eprf

\section{Concluding Remarks}

In this paper we have presented an isomorphism testing algorithm for
$n$-vertex chordal graphs of leafage $\ell$ which has running time
$t(\ell)\cdot n^{O(1)}$, where $t(\ell)$ is a double exponential
function not depending on $n$. A natural question is to improve the
running time dependence on the leafage.


The other problem of interest is isomorphism testing of \order{k}
hypergraphs for $b$-bounded color classes. Can we obtain an FPT
algorithm with both $k$ and $b$ as parameters, or with $k$ as
parameter for fixed $b$?\smallskip

\section{Acknowledgements}~~We thank the anonymous referees
of an earlier version for their valuable comments and corrections.
Roman Nedela was supported by GA\v{C}R~20-15576S.
Peter Zeman was supported by GA\v{C}R~20-15576S, GAUK~1224120, and by the Charles University project PRIMUS/21/SCI/014.

\end{document}